\documentstyle[epsfig]{mn}

\newcommand{\Msolar}{\mbox{\,$\rm M_{\odot}$}}        

\begin{document}

\title[X-ray variability in AGNs]{On X-ray variability in narrow line and broad line AGNs}
\author[Bian Weihao \& Zhao Yongheng]{W.Bian$^{1,2}$ and Y.Zhao$^2$\\
 $^{1}$Department of Physics, Nanjing Normal University, Nanjing
210097, China\\
$^{2}$National Astronomical Observatories, Chinese Academy of
 Sciences, Beijing 100012, China\\ }
\maketitle
\begin{abstract}
We assembled a sample of broad line and narrow lines AGNs observed
by ASCA, whose excess variances have been determined. The central
black hole masses in this sample can be obtained from the
reverberation mapping method and the width of H$\beta$ emission
line. Using the black hole masses and the bolometric luminosity,
the Eddington ratio can also be obtained. We confirmed the strong
anti-correlation between X-ray variability and the central black
mass founded by Lu \& Yu. We further found that narrow line
Seyfert 1 galaxies (NLS1s) also follow this relation with a larger
scatter and there is only a weak correlation between the X-ray
variability and the Eddington ratio, which suggest that rapid
variability and narrow lines in NLS1s are mainly due to small
central black holes in NLS1s, not the difference of circumnuclear
gas around NLS1s. A strong correlation was found between the hard
X-ray photon index and the Eddington ratio. If the suggestion of
two distinct accretion classes, namely ADAF and thin disk
accretion, in AGNs (Lu \& Yu) is correct, the strong photon
index-Eddington ration correlation showed that there exists a kind
of two zone accretion disk, in which the outer zone is a thin
disk, and the inner zone is an ADAF disk. Otherwise, the accretion
process is the thin disk accretion and the ADAF accretion is not
required.
\end{abstract}
\begin{keywords}
galaxies:active --- galaxies:nuclei --- X-rays: galaxies
\end{keywords}

\section{INTRODUCTION}
X-ray variability has long been known to be a common property of
active galactic nuclei (AGNs). There is not only long term but
also rapid X-ray variability. The rapid variability is usually
thought to come from the innermost region of the compact objects,
which is helpful in getting information about the central objects,
such as mass, accretion rate, geometry and radiation mechanisms
(Mushotzky et al. 1993).

Since the discovery of X-ray variability, there have been many
different quantities to describe the variability amplitude or the
timescale, such as the flux-doubling timescale (Barr \& Mushotzky
1986), normalized variability amplitude (Green et al. 1993), the
excess variance (Nandra et al. 1997), and the exponential
timescale (Bian \& Zhao 2003). Since then, a strong
anti-correlation between the variability and the luminosity
(hereafter variance-luminosity relation) was founded. Leighly
(1999) presented a X-ray variability analysis of a sample of 23
narrow line Seyfert 1 galaxies (NLS1s) observed by ASCA and found
that when NLS1 are included the variance-luminosity correlation of
AGN contains much more scatter than with BLS1 only, the excess
variance is typically an order of magnitude larger for NLS1s than
for Seyfert 1 with broad optical lines. Turner et al. (1999)
presented a sample of 36 Seyfert 1 galaxies observed by ASCA and
found a strong correlation between the excess variance and one
optical parameter, FWHM of H$\beta$, which is usually proposed to
be related with the "Eigenvector 1", the fundamental parameters of
the central engine (Boroson \& Green 1992). Lu \& Yu (2001) then
discussed the relation between the excess variance and the
reverberation central black hole masses (Kaspi et al. 2000)
(hereafter variance-mass relation) in a sample of 22 AGNs observed
by ASCA and found the relation between them to be strong. The
variance-mass correlation can provide a plausible explanation of
the variance-luminosity relation and its scatter (Turner et al.
1999; Leighly 1999; Ptak et al. 1998; Almaini et al. 2000).

Based on the circumnuclear gas difference in temperature, optical
depth, density, or geometry, Turner et al. (1999) provided an
attractive model, which does not depend on the difference in
central black mass, to explain differences of the excess variance
and the spectral index between NLS1s and Broad line AGNs (BL
AGNs). However Lu \& Yu (2001) suggested that the enhanced excess
variance in NLS1s is due to their central smaller black holes.
There are only a few NLS1s in the sample of Lu \& Yu (2001). It is
necessary to expand the sample to investigate the variance-mass
relation, especially to include more NLS1s.

Up to now, there are only 37 AGNs whose central black hole masses
are obtained by the reverberation mapping method (Ho 1998; Wandel
et al. 1999; Kaspi et al. 2000). Fortunately a relation between
the size of the broad line regions (BLRs) and the monochromatic
luminosity at 5100$\AA$ was founded by Kaspi et al. (2000). This
empirical formulae provides the estimation of the central black
hole masses for AGNs with available FWHM of H$\beta$. This kind of
mass estimation is widely used to investigate relations between
the black hole mass and radio luminosity (Woo \& Urry 2002), width
of O[III] line (Wang \& Lu 2001), and bulge mass (Bian \& Zhao
2003).

In this paper we use this empirical formula to estimate the
central black hole masses in an expanded sample of 41 AGNs
observed by ASCA and investigate the relations between the excess
variance, the photon index, the black hole mass, and the accretion
rate. In particular, we wanted to determine whether the
variance-mass relation founded by Lu \& Yu (2001) would alter when
we included more AGNs, especially NLS1s. The sample is described
in Sect. 2 and the central black hole mass and the Eddington ratio
are calculated in Sect. 3. The statistical analysis and discussion
are presented in Sect. 4. Finally we summarize our conclusions in
Sect. 5. All cosmological calculations in this paper assume
$H_{0}=75 km s^ {-1}, \Omega =1.0, \Lambda=0$.

\section{SAMPLE}

The sample used in Turner et al. (1999) consists of 36 Seyfert 1
galaxies available from the ASCA archive up to Nov. 1998. The
criteria in Turner et al. (1999) is that only the objects whose
light curves have at least 20 counts per 256-second time bin and
at least 20 bins are included in their sample. They used the
excess variance to describe the X-ray variability. If one
designates the count rates for the N points in each light curve as
$X_{i}$, with errors $\sigma_{i}$, the X-ray excess variance is
defined by,
$$ \sigma_{rms}^2 = \frac{1}{N\mu^2} \sum_{i=1}^N
[(X_i-\mu)^2-\sigma^2_i]\ ,
$$
The error on $\sigma_{rms}^2$ is $s_D/(\mu^2\sqrt{N})$, where
$$
s^2_D=\frac{1}{N-1} \sum_{i=1}^N
{\{[(X_i-\mu)^2-\sigma^2_i]-\sigma_{rms}^2\mu^2\}}^2\
$$
which is only the statistical error (Turner et al. 1999).

Lu \& Yu (2001) also searched the ASCA archive up to Oct. 1999 to
do a timing analysis using the same criteria as Turner et al.
(1999). Here we adopt the variance data from Turner et al. (1999)
and Lu \& Yu (2001).

As in Turner's paper (1999), for convenience we make such a split
here by referring to objects with FWHM H $\beta < 2000$ km/s as
narrow line AGNs (NL AGNs) and objects with FWHM H $\beta
> 2000$ km/s as broad line objects (BL AGNs). At last, our sample
consists of 41 AGNs, in which there are 18 NL AGNs and 23 BL AGNs.
Although this sample is not a complete one, we suggest this large
assembled sample provides new information about the difference
between BL AGNs and NL AGNs. For objects with many values of the
excess variance and the photon index, we adopt their mean values.
The standard error of the mean is so small that we adopt the error
from the above formula.

\section{BLACK HOLE MASS AND EDDINGTON RATIO}
\subsection{Estimation of black hole masses}
There are 22 AGNs in our sample whose central black hole masses
are estimated by the reverberation mapping method. We use the
black hole masses tabulated by Kaspi et al. (2000); values for
three AGNs (Mrk 279, NGC 3516, and NGC 4593) were taken from Ho
(1998). At the same time, we also use the B-magnitude and FWHM of
H$\beta$ to estimate the masses for all 41 objects in our sample.
We estimated the size of the BLR using the empirical correlation
between the size and the monochromatic luminosity at
5100$\rm{\AA}$ (Kaspi et al. 2000):
\begin{equation}
\rm{R_{BLR}=32.9(\frac{\lambda L_{\lambda}(5100 \rm{\AA})}{10^{44}
erg \cdot s^{-1}})^{0.7} ~~\rm{lt-days}} \label{eqn1}
\end{equation}
where $\lambda L_{\lambda}(5100 \rm{\AA)}$ is estimated from the
B-magnitude by adopting an average optical spectral index of -0.3
and accounting for Galactic redding and K-correction (Veron-Cetty
et al. 2001). If the H$\beta$ widths reflect the Keplerian
velocity of the line-emitting BLR material around the central
black hole, then the viral mass is given by:
\begin{equation}
\rm{M_{bh}=R_{BLR}V^{2}G^{-1}} \label{eqn2}
\end{equation}
where G is the gravitational constant, V is the velocity of the
line-emitting material. V can be estimated from the H$\beta$
width. Assuming the random orbits, Kaspi (2000) related the $V$ to
the FWHM of the H$\beta$ emission line by $V=(\sqrt{3}/2) \rm
FWHM_{[H\beta]}$.

\subsection{Eddington ratio}
In order to investigate the relation between X-ray variability and
the accretion rate, the Eddington rates are calculated for these
41 AGNs. We calculated the ratio of the bolometric luminosity
$L_{bol}$ to the Eddington luminosity $L_{Edd}$. $L_{bol}$ is
usually calculated by $L_{bol}=9\lambda L_{\lambda}(5100
\AA)$(Kaspi et al. 2000), where $L_{\lambda}(5100 \AA)$ is the
monochromatic luminosity at 5100$\rm{\AA}$. The value of
$L_{bol}/L_{Edd}$ gives a direct measurement of the Eddington
ratio.

The estimated black hole masses are listed in table 1. Column (1)
lists the name of objects, column (2) the FWHM of H$\beta$ $(km
s^{-1})$, Column (3) the log of black hole mass in $\Msolar$
calculated using equations (1) and (2), Column (4) the log of the
reverberation mapping mass in $\Msolar$ adopted from Kaspi et al.
(2000) and Ho (1998) , and Column (5) the Eddington ratio. The
excess variances are from Table 1 in Turner et al. (1999) and
Table 1 in Lu \& Yu (2001). The photon indices $\Gamma_{ASCA}$ are
from Table 1 in Turner et al. (1999). For objects with both the
reverberation mapping mass and the calculated mass using equations
(1) and (2), we adopt the former because the former is more
reliable.

\section{RESULTS AND DISCUSSION}

\begin{figure}
\begin{center}
\centerline{\epsfig{figure=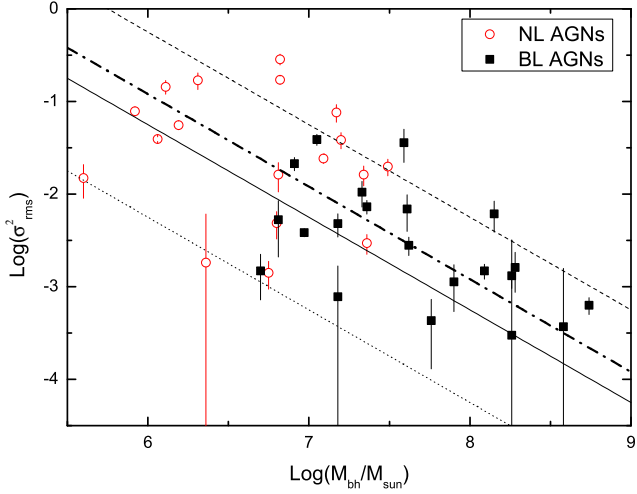,width=10cm}} \caption{Excess
variance in the 0.5-10keV band versus the central black hole mass.
The filled squares represent the BL AGNs; the open circles
represent NLS1s. The dash-dot line is our best fit.}
\end{center}
\end{figure}

\begin{figure}
\begin{center}
\centerline{\epsfig{figure=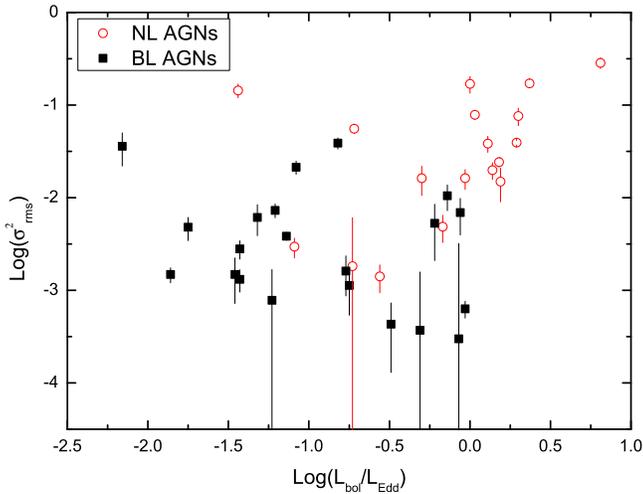,width=10cm}} \caption{Excess
variance in the 0.5-10keV band versus the Eddington ratio. The
denotation is the same as that in Fig. 1.}
\end{center}
\end{figure}

\subsection{Variance-mass relation}
In Fig. 1 we show the excess variance versus the central black
hole masses. In Fig. 2 we show the excess variance versus the
Eddington ratio. We find the tight variance-mass correlation
founded by Lu \& Yu (2001) still remains when we included many
other AGNs, especially NL AGNs. In Fig. 1 we also show the
straight lines plotted in Fig. 1 in Lu \& Yu (2001). Most of the
objects are located in the region between the lines
$log\sigma_{rms}^{2} = 3.75-log(M_{bh}/\Msolar)$ and
$log\sigma_{rms}^{2} = 5.75-log(M_{bh}/\Msolar)$. We fix the slope
of the variance-mass relation as -1 to fit the data in Fig.1. A
Spearman rank test gives $log\sigma_{rms}^{2} =
(5.08\pm0.11)-log(M_{bh}/\Msolar)$ and the correlation
coefficients (R) is -0.65 corresponding to a probability of
$P<10^{-4}$ that the correlation is caused by a random factor.

There are astrophysical reasons to relate the excess variance with
the accretion rates. However we find there is no strong
correlation between them. We plot the excess variance versus the
Eddington ratio in Fig. 2. A Spearman rank test gives
$log\sigma_{rms}^{2} = (-1.88\pm 0.16)+(0.39\pm
0.17)log(L_{bol}/L_{Edd})$ ($R=0.35$, $P=0.0309$). When
considering only NL AGNs or BL AGNs, we find there is no
correlation between the excess variance and the Eddington ratio if
we exclude NGC 3227 (Lu \& Yu 2001).

\subsection{Relation between photon index and Eddington ratio}
Although weak relation between the excess variance and the
Eddington ratio is found here, we find there is a strong
correlation between the photon index in 2-10keV ($\Gamma_{ASCA}$)
and the Eddington ratio (Fig. 3). A Spearman rank test gives
$\Gamma_{ASCA} = (2.11\pm 0.05)+(0.26\pm
0.05)log(L_{bol}/L_{Edd})$ ($R=0.70$, $P<10^{-4}$). The
correlation between $\Gamma_{ASCA}$ and Eddington ratio still
remains medium strong if we only consider BL AGNS (R=0.26) or NL
AGNs (R=0.48).

\begin{figure}
\begin{center}
\centerline{\epsfig{figure=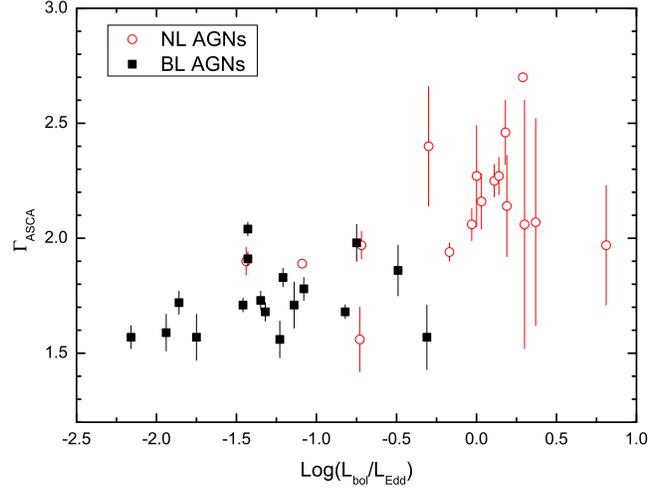,width=10cm}}

\caption{The photon index versus the Eddington ratio. The
denotation is the same as that in Fig. 1.}
\end{center}
\end{figure}

\subsection{Discussion}
We use equations (1) and (2) to calculate the black hole masses
for all 41 AGNs in our sample. The errors of the calculated black
hole masses using equations (1) and (2) is mainly from the
accuracy of equation (1); the geometry and the dynamics of the
BLRs, especially the inclination to the line of sight in NL AGNs
(Bian \& Zhao 2002). The error in the mass estimation using
equations (1) and (2) is about 0.5 dex (Wang \& Lu 2001). Fig. 4
shows that our calculated masses from equations (1) and (2) is
consistent with the reverberation mapping masses (see Table 1).
The variance-mass relation in NL AGNs is consistent with that in
BL AGNs founded by Lu \& Yu (2001). It is plausible that there
exists a variance-mass relation for not only BL AGNs but also NL
AGNs. There is a weak correlation between the excess variance and
the Eddington ratio. There is no correlation between them if we
just consider BL AGNs or NL AGNs. Turner et al. (1999) have
suggested that the different circumnuclear gas of NL AGNs may lead
to the different X-ray variance between BL AGNs and NL AGNs. The
difference of the circumnuclear gas is caused by the higher
accretion rate in NL AGNs. The strong variance-mass correlation
and the weak correlation between the variability and the Eddington
ratio suggest that the enhanced hard X-ray excess variance in NL
AGNs founded by Leighly (1999) is mainly due to the smaller black
hole, not the difference of circumnuclear gas around NL AGNs. The
variance-mass anti-correlation can be interpreted that X-ray
variance is a result of some global coherent variations in the
X-ray emission region which is scaled by the size of the black
hole in AGNs (Lu \& Yu 2001).

Equation (1), used to estimate the black hole mass, is derived in
the sample of Kaspi et al. (2000). However, most AGNs in the
sample of Kaspi et al. (2000) are BL AGNs. Whether equation (1) is
suitable for NL AGNs is a matter of debate. The masses for NLS1s
estimated from equation (1) appear to be consistent with that from
the the mass-velocity dispersion relation in AGNs (Wang \& Lu
2000), which showed that equation (1) can be used in NL AGNs in
spite of the larger uncertainty. Turner et al. (1999) showed that
there is a strong correlation between the excess variance and the
FWHM H$\beta$. For our sample we also find a strong variance-FWHM
correlation ($R=-0.64$, $P<10^{-4}$), which is scatter than the
variance-mass correlation ($R=-0.65$, $P<10^{-4}$). The stronger
variance-mass correlation showed that FWHM H$\beta$ can be
combined with the luminosity to estimate the central mass. Perhaps
the early finding of the variance-FWHM correlation is due to the
variance-mass correlation. At the same time we should notice the
uncertainty in equation (1) leads to just a little stronger
variance-mass correlation.

\begin{figure}
\begin{center}
\centerline{\epsfig{figure=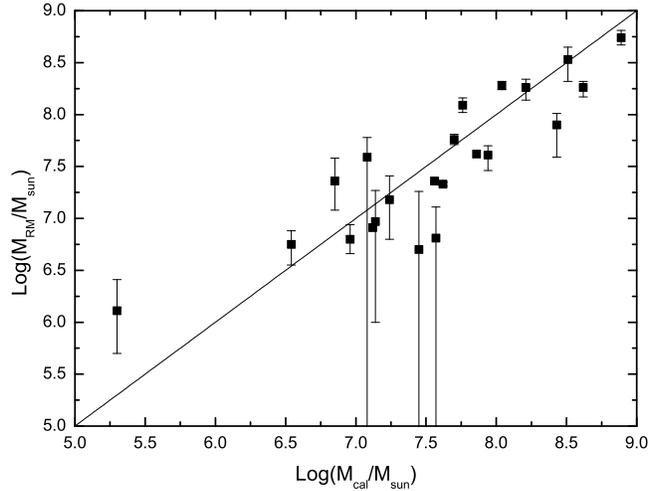,width=10cm}}

\caption{T he reverberation mapping masses versus the black hole
masses calculated from equations (1) and (2). The solid line shows
1:1.}
\end{center}
\end{figure}

We also find there exists a strong correlation between the photon
index $\Gamma_{ASCA}$ and the Eddington ratio. Lu \& Yu (1999)
compiled a sample of Seyfert 1 galaxies and QSOs and found these
objects to be distributed in two distinct classes in the plane of
ROSAT soft X-ray photon index versus the Eddington ratio (the
ratio of the ionizing luminosity to the Eddington luminosity).
There is a strong correlation between ROSAT photon index and the
Eddington ratio in these two classes. They also plot
$\Gamma_{ASCA}$ versus the Eddington ratio for several objects
with available ASCA data. Here we use the bolometric luminosity
rather than the ionizing luminosity to calculate the Eddington
ratio and find $\Gamma_{ASCA}$ is strongly correlated with the
Eddington ratio. The enhanced $\Gamma_{ASCA}$ in NL AGNs (Leighly
1999) is intimately related with the Eddington ratio. First we
assume that the interpretation of Lu \& Yu (1999) on the split in
the photon index-accretion rates plane is correct, there are two
distinct accretion disks in Seyfert 1 galaxies and QSOs. If we
take $log(L_{bol}/L_{Edd})=-1.1$ (Lu \& Yu 1999) as the critical
value for ADAF (advection-dominated accretion flow) disk class and
thin disk class, from Fig. 3 we find the $\Gamma_{ASCA}$ increases
with the Eddington ratio in ADAF and thin disk classes. The
relation between $\Gamma_{ASCA}$ and the Eddington ratio can be
understood in the frame of the accretion disk and the corona. For
a thin disk, the disk luminosity to irradiate the corona increases
as the Eddington rate increases. This can cause the corona to cool
efficiently owing to Compton cooling and cooler corona producing
few hard X-ray photons leads to large $\Gamma_{ASCA}$. For the
ADAF model of low accretion rate, the optical depth increases and
causes a correspondent increase in the Compton $\gamma$-parameter
when the Eddington ratio increases (Esin et al. 1997). Larger
optical depth will result in a harder and smoother X-ray spectrum
index, which is in conflict with our founded relation between
$\Gamma_{ASCA}$ and the Eddington ratio. A accretion disk
consisting two zones (outer thin disk and inner ADAF disk) is our
preferable interpretation. The truncation radius of the two zones
decreases with increasing Eddington ratio. The ADAF X-ray photon
index $\Gamma_{ASCA}$ becomes dramatically softer because the
radiation from the disk is Compton-scattered by the hot gas in
ADAF as the Eddington rate increases. If the interpretation of Lu
\& Yu (1999) is not correct, the strong correlation between
$\Gamma_{ASCA}$ and the Eddington ratio in the sample of NL and BL
AGNs suggests the ADAF disk is not required and there exists a
simple thin accretion disk. Much work on detail spectral fitting
of AGNs using ADAF and/or thin disk models is needed to clarify
this question in the future.

\section{CONCLUSIONS}
We investigated the relations between hard X-ray variability,
photon index, the black hole mass, and the Eddington ratio in a
collected sample of 41 BL AGNs and NL AGNs. NL AGNs follow the
same variance-mass relation in BL AGNs founded by Lu \& Yu (2001).
The X-ray variability is mainly due to the black hole mass, not
the accretion rate. A strong correlation between the hard X-ray
photon index and the Eddington ratio suggests there exists the
thin disk in the inner region of the AGNs. If the suggestion of
two distinct accretion classes (Lu \& Yu 1999) is correct, there
exists a kind of two zone accretion disk, in which the outer zone
is a thin disk, and the inner zone is an ADAF disk. Otherwise, the
accretion process is the thin disk accretion and the ADAF
accretion is not required.

\section*{ACKNOWLEDGMENTS}
This work has been supported by the NSFC (No. 10273007).

\begin{table}
\begin{tabular}{lcccc}
\hline \hline

Name & FWHM & $M_{cal}$ & $M_{RM}$ & $\dot m$ \\
\hline
Mrk335       &  1620 &   6.96   &    6.80  &  -0.17        \\
IZW1         &  1240 &   7.20   &    --      &   0.11        \\
TonS180      &  1120 &   7.09   &    --      &   0.18        \\
Fairall9     &  5780 &   8.43   &    7.90  &  -0.75        \\
RXJ0148-27   &  1050 &   7.17   &     --     &   0.30        \\
Nab0205+024  &  1330 &   7.49   &     --     &   0.14        \\
Mrk1040      &  1830 &   6.36   &     --     &  -0.73        \\
LB1727       &  2800 &   8.58   &     --     &  -0.31        \\
3C120        &  1910 &   6.85   &    7.36  &  -1.09        \\
Ark120       &  5800 &   8.21   &    8.26  &  -1.43        \\
MCG+8-11-11  &  3630 &   7.18   &     --     &  -1.23        \\
H0707-495    &  1000 &   6.31   &     --     &   0.00   \\
PG0804+761   &  2757 &   8.04   &    8.28  &  -0.77        \\
PG0844+349   &  2210 &   7.62   &    7.33  &  -0.14        \\
Mark110      &  1430 &   6.54   &    6.75  &  -0.56        \\
PG0953+415   &  3110 &   8.62   &    8.26  &  -0.07        \\
NGC3227      &  4920 &   7.08   &    7.59  &  -2.16        \\
RE1034+396   &  1500 &   6.81   &     --     &  -0.30        \\
NGC3516      &  4760 &   7.56   &    7.36$^{a}$  &  -1.21        \\
NGC3783      &  3790 &   7.14   &    6.97  &  -1.14        \\
Mark42       &   670 &   5.60   &     --     &   0.19        \\
NGC4051      &  1170 &   5.30   &    6.11  &  -1.44        \\
NGC4151      &  5910 &   7.24   &    7.18  &  -1.75        \\
PG1211+243   &  1832 &   7.71   &    7.61  &  -0.06        \\
Mrk766       &   850 &   5.92   &     --     &   0.03        \\
3C273(PG1226) &  2810 &   8.89   &    8.74  &  -0.03        \\
NGC4593      &  3720 &   7.12   &    6.91$^{a}$  &  -1.08        \\
IRAS13224-38 &   620 &   6.82   &     --     &   0.81        \\
MCG-6-30-15  &  1700 &   6.19   &      --    &  -0.72        \\
IC4329A      &  4800 &   7.45   &    6.70  &  -1.46        \\
Mrk279       &  5360 &   7.86   &    7.62$^{a}$  &  -1.43        \\
PG1404+226   &   880 &   6.82   &     --     &   0.37        \\
NGC5548      &  5610 &   7.76   &    8.09  &  -1.86        \\
Mrk478       &  1450 &   7.34   &     --     &  -0.03        \\
Mrk841       &  5470 &   8.15   &     --     &  -1.32        \\
Mrk290       &  2500 &   7.05   &     --     &  -0.82        \\
3C390.3      & 10000 &   8.51   &    8.53  &  -1.94        \\
Mrk509       &  2270 &   7.70   &    7.76  &  -0.49        \\
Ark564       &   720 &   6.06   &    --      &   0.29        \\
NGC7469      &  3388 &   7.57   &    6.81  &  -0.22        \\
MCG-2-58-22  &  6360 &   8.54   &    --      &  -1.35        \\

\hline
\end{tabular}
\caption{The central black hole mass and the Eddington
ratio for broad line AGNs and narrow line AGNs. The reverberation
mapping masses are adopted from Kaspi et al.(2000), except for
those objects labelled with a, which are adopted from Ho (1998).}
\end{table}

\end{document}